\documentclass{jpsj-suppl}

\usepackage{txfonts}
\usepackage{bm}
\usepackage{eurosym}
\usepackage{amsmath}
\usepackage{graphicx}
\usepackage{siunitx}
\usepackage{amsmath}
\usepackage[greek,english]{babel}

\newcommand{\norm}[1]{\left\lVert#1\right\rVert}

\title{$\bar{\textrm{H}}^{+}$ Sympathetic Cooling Simulations with a Variable Time Step}

\author{Nicolas Sillitoe$^{1}$, Jean-Philippe Karr$^{1,2}$, Johannes Heinrich$^{1}$, Thomas Louvradoux$^{1}$, Albane Douillet$^{1,2}$, Laurent Hilico$^{1,2}$}

\inst{$^{1}$ Laboratoire Kastler Brossel, UPMC-Sorbonne Universit\'es, CNRS, ENS-PSL Research University, Coll\`ege de France \\
$^{2}$Dpt de Physique, Universit\'e d'Evry Val d'Essonne, 91025 EVRY France}

\email{laurent.hilico@lkb.upmc.fr}

\recdate{June 16, 2016}

\abst{ In this paper we present a new variable time step criterion for the velocity-Verlet algorithm allowing to correctly simulate the dynamics of charged particles exchanging energy via Coulomb collisions while minimising simulation time. We present physical arguments supporting the use of the criterion along with numerical results proving its validity. We numerically show that $\bar{\textrm{H}}^{+}$ ions with 18~meV initial energy can be captured and sympathetically cooled by a Coulomb crystal of $\textrm{Be}^{+}$ and $\textrm{HD}^+$ in less than 10~ms, an important result for the GBAR project.
}

\kword{Sympathetic Cooling}

\begin{document}
\maketitle

\section{Introduction}
A Paul Trap allows very long trapping times for ions, which in combination with cooling leads to applications
in fields such as high resolution spectroscopy~\cite{Koelemeij2007,Biesheuvel2016}, quantum computation,
quantum simulations~\cite{Blatt2012} and cold chemistry~\cite{Eberle2015,Willitsch2012}. Some ions such as $^{9}\textrm{Be}^{+}$ can be conveniently laser cooled~\cite{Wineland1978,Nauhauser1978},
most, however, cannot. One way to overcome this is sympathetic cooling
whereby instead of trapping only the species of interest another species,
which can be laser cooled, is simultaneously trapped. The species which cannot be
laser cooled will thermalise via Coulomb interaction with the other
species, forming a cold two-component Coulomb crystal with a temperature bounded
by the Doppler cooling limit, e.g. 0.47~mK or 60~neV in the case of Be$^+$.
Sometimes the ions cannot be created in situ and therefore have to be externally loaded at relatively high energies of 0.1-10~eV. Such is the case of highly charged ions~\cite{Schmoger2015} and antimatter ions~\cite{Perez2015}  that are (or will be) created in dedicated sources and are of interest for fundamental physics experiments. One example is the GBAR experiment which aims to cool $\bar{\textrm{H}}^{+}$ ions made of an antiproton and two positrons at CERN and study their free fall to measure the gravity constant $\bar{g}$ on antimatter~\cite{Indelicato2014,Walz2004,Perez2012}.

One crucial step of the GBAR project is the capture and sympathetic cooling of
$\bar{\textrm{H}}^{+}$ ions, so it is important to accurately evaluate the sympathetic cooling time by a laser cooled Be$^+$ ion crystal. Sympathetic cooling of externally loaded ions has been recently achieved for Ar$^{13+}$~\cite{Schmoger2015,Schmoger2015b} but the dynamics of the process have been little studied so far~\cite{Bussmann2006} especially for the case of very different mass-to-charge ratios as is the case in GBAR (9:1)~\cite{Hilico2014}.
Our goal is to numerically study this Doppler-cooling step.
In Sec.~\ref{section_model}, we briefly introduce the numerical model used
for the simulations. In Sec.~\ref{section_time_step}, we discuss the choice of the simulation time step and propose a new scheme to well describe Coulomb interactions that lead to sympathetic cooling. In Sec.~\ref{section_result}, we discuss our first numerical results showing that $\bar{\textrm{H}}^{+}$ can be cooled by a laser cooled Be$^+$/HD$^+$ ion crystal.

\section{Ion crystal dynamics model}
\label{section_model}
\begin{figure}[h]
\begin{center}
\includegraphics[scale=0.65]{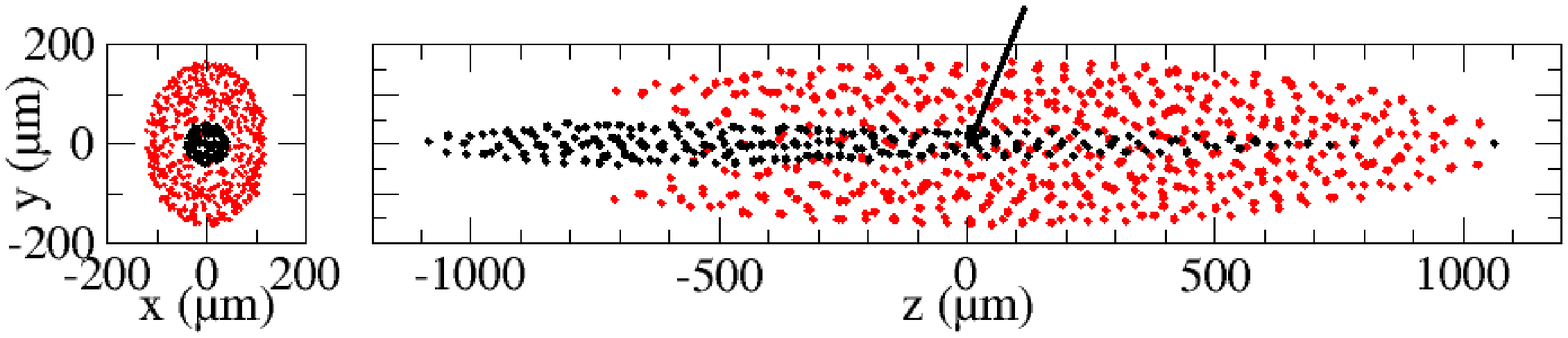}
\caption{Projection in the xy (left) and zy (right) planes of a simulated ion crystal made of 500 Be$^+$ and 200 HD$^+$. The laser cooling beam is applied towards the increasing z direction, pushing the Be$^+$ ions in that direction. The HD$^+$ ions are more likely located on the  other side of the cloud. The arrow indicates the location of the $\bar{\textrm{H}}^+$ ion at the end of the simulation shown in Fig.~\ref{Fig_trajectories}. }
\label{Fig_ion_cloud}
\end{center}
\end{figure}

For the GBAR experiment, the idea~\cite {Walz2004} is to sympathetically cool a high velocity $\bar{\textrm{H}}^+$ ion using a laser cooled trapped Be$^+$ ion crystal.
In a linear RF Paul trap, a laser cooled ion crystal in the Coulomb crystal regime has an ellipsoidal shape~\cite{Turner1987} as shown in Fig~\ref{Fig_ion_cloud}. The inter-ion distance is a few tens of microns and for ion numbers $N$ of a few thousand the crystal typically has dimensions of a few millimeters. We therefore have a low density mesoscopic system which would be poorly described by mean-field methods so we cannot use $NlogN$ approximate methods~\cite{Chapman2010} for the Coulomb interaction and we have to exactly compute it using the $N^{2}$ all-pairs approach.
We solve Newton's equations including the time-dependent confinement electric fields of the linear trap, the exact Coulomb force between all ion pairs and the laser interaction~\cite{Zhang2007,Schiller2003,Okada2010,Marciante2010,Marciante2011,Marciante2012}.

The confinement field derives from the potential~\cite{Hilico2014}
\begin{equation}
V(x,y,z,t)=(U_0+V_{\rm RF}\cos(\Omega t))\frac{x^2-y^2}{2r_0^2}+\frac{1}{2}m_i\omega_{i,z}^2 (z^2-(x^2+y^2)/2),\label{eq_RF_potential}
\end{equation}
including the radial confinement in the $x$ and $y$ directions. For this study, we use $U_0$=0.1~V ,
$V_{\rm RF}$=100~V, $r_0$=3.5~mm and $\Omega$=2$\pi\ \times$ 13~MHz.
The longitudinal confinement in the $z$ direction is described in terms of the axial
oscillation frequency $\omega_z$=2$\pi\times$100~kHz for
$\bar{\textrm{H}}^+$ (which is smaller by a factor of $\sqrt{3}$ and 3 for HD$^+$ and $^9$Be$^+$ respectively).

We describe the interaction with the cooling laser as a stochastic process of absorption (depending on the ion's velocity through Doppler effect) spontaneous and stimulated emission, adding the corresponding velocity kicks to the laser cooled ions~\cite{Pwctp2016,Rouse2015}.
The laser cooling beam is aligned with the trap axis. The waist of 1~mm is located at the trap center. The laser detuning is $-\Gamma$ where $\Gamma$=19~MHz is the natural width of the Be$^{+}$ cooling transition and laser intensity is 1.5 times the saturation intensity.

The ion trajectories are computed using the velocity-Verlet algorithm~\cite{Verlet1967} given by
\begin{equation}
\left\lbrace
\begin{array}{c}
x_i\\
a_i\\
v_i
\end{array}
\right.
\rightarrow
\left\lbrace
\begin{array}{l}
x_{i+1}=x_i+v_i\ \delta t+\frac{a_i\delta t^2}{2}\\
a_{i+1}=F(x_{i+1},t_{i+1})\\
v_{i+1}=v_i+\frac{a_i+a_{i+1}}{2}\ \delta t
\end{array}
\right. \label{NS-eq-leapfrog}
\end{equation}
where $x_i=x(t_i)$, $v_i=v(t_i)$, $a_i=a(t_i)$, $F(x_i,t_i)$ is the force at time $t_i$ and $\delta t$ is the integration time step.

\section{Choice of the integration time step}\label{section_time_step}
To accurately describe the Radio Frequency trapping potential (RF) of the linear Paul trap the integration time step $\delta t$ needs to verify
\begin{equation}
\delta t << 2\pi/\Omega
\label{RF Criterion}
\end{equation}
As we will show, the proper description of coulomb interactions between ions can lead to time steps orders of magnitude smaller than what Eq.~(\ref{RF Criterion}) prescribes. Eq.~(\ref{RF Criterion}) therefore gives an upper bound on the time step that can be used. We have checked that setting it to \SI{0.1}{\nano\second} (about $1/769$ of the RF period) in our simulations, the trajectory of a single ion in the RF field is converged for simulations longer than 10~ms.
The description of laser cooling in terms of random absorption and emission events imposes time steps much longer than the optical period, otherwise one would have to describe the laser interaction in terms of Bloch equations as discussed in~\cite{Rouse2015}. For Be$^+$ cooling at 313~nm, the optical period is 1~fs, so the time step can be adapted over several orders of magnitude.

\subsection{Coulomb interaction simulations with a fixed time step}
\label{fixed time step section}
\begin{figure}[h]
\begin{center}
\includegraphics[scale=0.5]{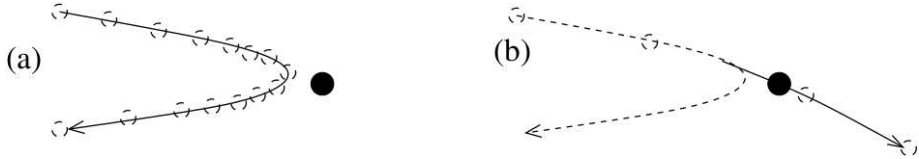}
\end{center}
\caption{Coulomb collision between two ions in the frame of the target ion. (a) The time step is short enough and the collision is well described. (b) The time step is too long and the ions go through each other instead of repelling each other.}
\label{Time step Problem Figure}
\end{figure}

In Figure~\ref{Time step Problem Figure} we consider a Coulomb collision between two particles and illustrate that if the time step is too long, a Coulomb collision between two ions may be so poorly described that the two ions go through each other instead of repelling each other.
To understand the time step requirements of simulating the Coulomb interaction we simulated head-on 1D Coulomb collisions of two ions of equal masses and initially separated by 1~mm using a constant time step velocity-Verlet algorithm in the absence of the trapping field and the laser interaction. We send one ion at a given energy onto the other ion at rest. This problem has an analytical solution which predicts that the projectile ion transfers all its kinetic energy to the second ion.
\begin{figure} [h]
\begin{center}
\includegraphics[scale=0.275]{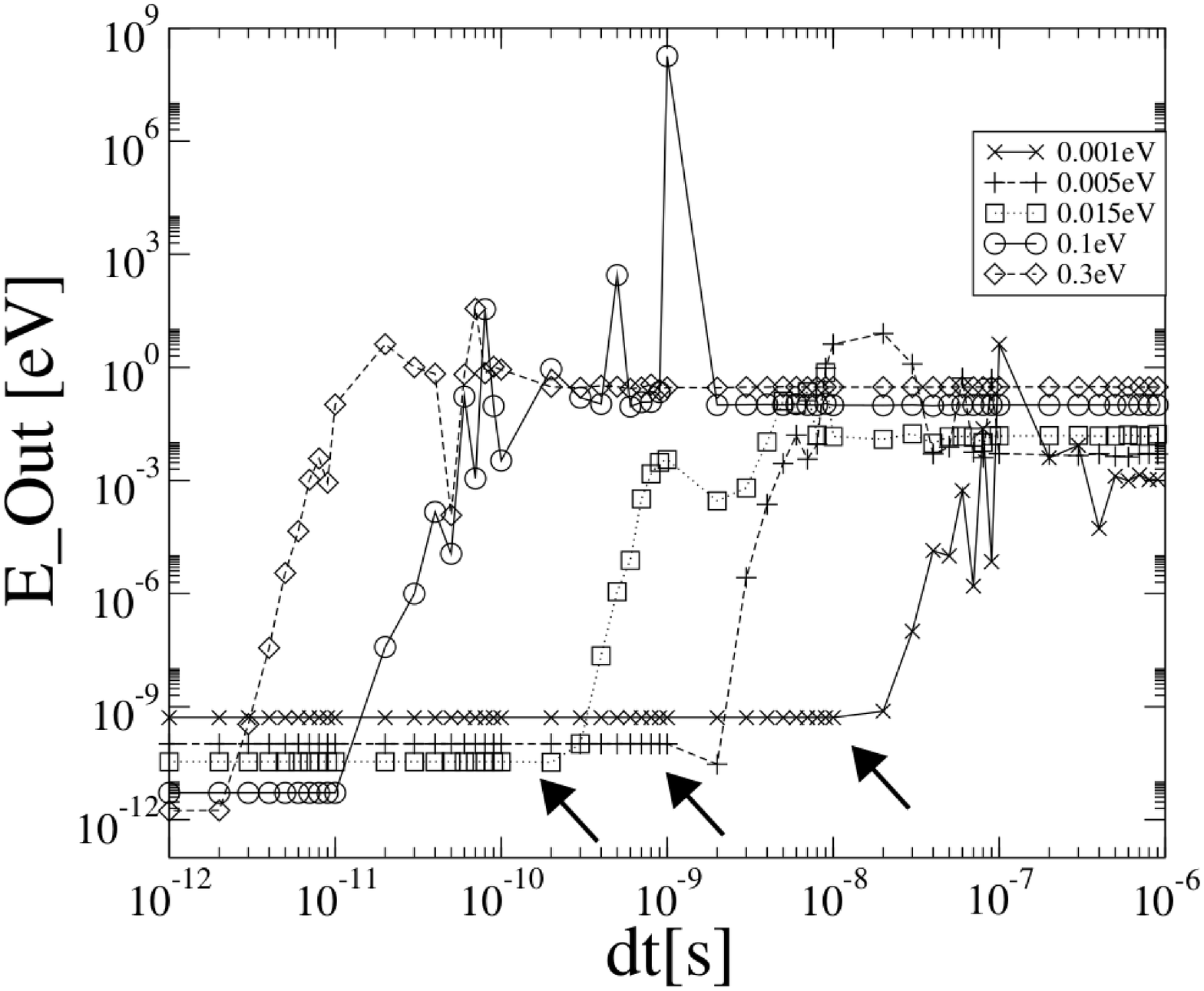}
\includegraphics[scale=0.275]{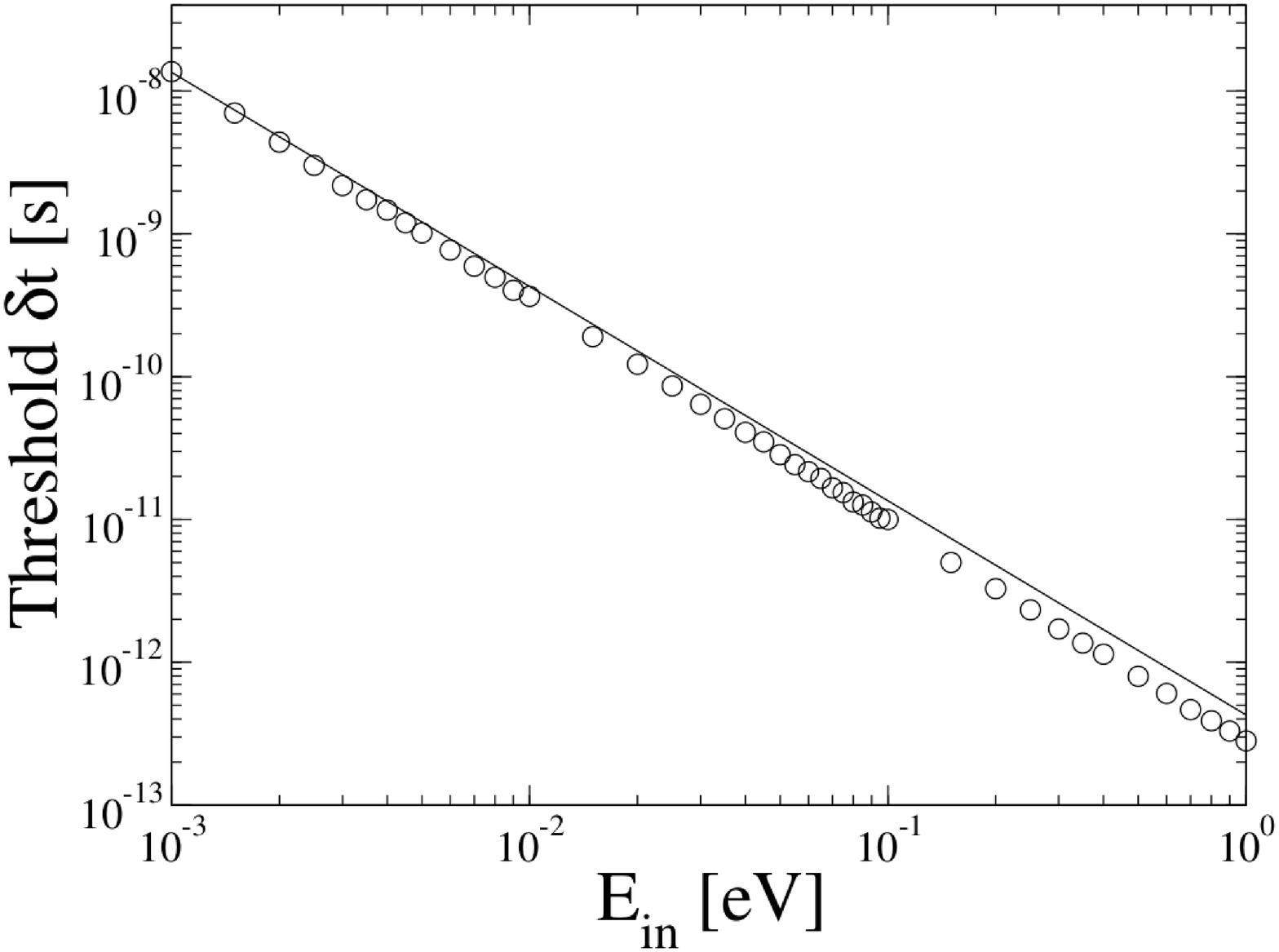}
\end{center}
\caption{Left: Outgoing energy of a projectile ion after colliding with a stationary ion of equal mass \SI{9}{\atomicmassunit} starting \SI{1}{\milli\meter} away from each other, for different projectile ion energies, versus the constant time step used to simulate the collision. Arrows indicate the threshold time step below which the collision is well described. Right: The open circles show  the threshold time step versus projectile initial kinetic energy, solid line is a fit to $\delta t=cE_{in}^{-\frac{3}{2}}$.}
\label{E_Out Figure}
\end{figure}

In Fig.~\ref{E_Out Figure} we show the energy of the projectile ion after the collision versus time step for different projectile energies.
At a given projectile energy we can see that for a too long time step the ions don't exchange much energy. However, there is a threshold time step below which we can reproduce the expected result of the projectile ion losing all its energy. Notice the intermediate regime where the time step is close to being small enough, the outgoing energy of the projectile ion can fluctuate quite wildly as the ions can come closer than their minimum approach distance, numerically adding energy to the system.

We can interpret the threshold time step by saying that in a time step the ions should move much less than their minimum approach distance. At the beginning of the collision most of the mechanical energy is in the kinetic energy $E_{in}$ of the projectile ion going at speed $v_{in}$ in the lab reference frame. The minimum approach distance $d_{min}$ is therefore given by
\begin{equation}
d_{min}=\frac{q_{1}q_{2}}{2\pi\epsilon_{0}\mu v_{in}^{2}}=\frac{q_{1}q_{2}}{\pi\epsilon_{0}m v_{in}^{2}}
\end{equation}
with $q_{1}$, $q_{2}$ the charges of the two ions and $\mu=m/2$ the reduced mass.
The displacement should obey $\delta r\approx v\ \delta t << d_{min}$ leading to
\begin{equation}
\delta t << \frac{q_{1}q_{2}}{\pi\epsilon_{0}m v_{in}^{3}}=\frac{q_{1}q_{2}\sqrt{m}}{\sqrt{8}\pi\epsilon_{0}}E_{in}^{-\frac{3}{2}}
\label{Constant time step interpretation law}
\end{equation}
where we have obtained the right hand side of the equation by upper bounding the relative speed of the two ions $v$ by $v_{in}$. We have fitted the threshold time steps found by such simulations to $\delta t=cE_{in}^{-\frac{3}{2}}$ and have found excellent agreement, finding that $c$ should be approximately 4 times smaller than the right hand side of Eq.~(\ref{Constant time step interpretation law}) for correct simulations. Figure~\ref{E_Out Figure} shows that the threshold time step is as small as \SI{1}{\pico\second} for only \SI{0.3}{\electronvolt} of initial kinetic energy. This makes sympathetic cooling simulations of high temperature particles with a constant time step extremely demanding.

\subsection{Variable time step criterion}
\label{Variable Time Step section}
Realistic dynamics of an ion crystal may involve fast ions and require very short time steps, e.g. if collisions with neutrals or exothermic reactions take place or as in the present case, if a fast ion is injected in a cold ion crystal to be sympathetically cooled.
In this section, we show that a variable time step scheme allows us to have much longer time steps on average while accurately describing Coulomb interactions.

From the ideas of Sec.~\ref{fixed time step section}, we say that at every time step, for every ion pair, the relative change in distance $\delta d_{ij}$ should obey
\begin{equation}
\delta d_{ij} << d_{ij}
\label{d Criterion}
\end{equation}
with $d_{ij}$ the distance between the two ions. We can reformulate the inequality in Eq.~(\ref{d Criterion}) as
\begin{equation}
\delta d_{ij}=\alpha d_{ij}
\label{d Criterion 2}
\end{equation}
with $\alpha$ a constant to be chosen such that $\alpha << 1$. For velocity-Verlet integration position updates are given by
\begin{equation}
\delta \bm{r_i}=\bm{v_i}\delta t+\frac{1}{2}\bm{a_i}\delta t^{2}\
\end{equation}
therefore the displacement $d_{ij}$ between two ions is upper bounded by
\begin{equation}
\delta d_{ij}\leq\norm{\bm{v_{ij}}}\delta t+\frac{1}{2}\norm{\bm{a_{ij}}}\delta t^{2}
\label{Verlet Position Increment}
\end{equation}
with $\bm{v_{ij}}$ and $\bm{a_{ij}}$ the relative velocity and acceleration.
Inserting Eq.~(\ref{Verlet Position Increment}) in Eq.~(\ref{d Criterion 2}) and solving the second degree equation for $\delta t$ one finds the following time step
\begin{equation}
\delta t_{ij}=\frac{-\norm{\bm{v_{ij}}}+\sqrt{\norm{\bm{v_{ij}}}^{2}+2\norm{\bm{a_{ij}}}\alpha\ d_{ij}}}{\norm{\bm{a_{ij}}}}.
\label{d Criterion dt}
\end{equation}
Eq.~(\ref{d Criterion dt}) is specific to the case of a Velocity Verlet integration. Indeed, we have found that approximating $\delta d_{ij}$ by $\norm{\bm{v_{ij}}}\delta t$ wasn't sufficient to describe a Coulomb interaction because as the ions approach their minimal approach distance and as the velocity $\norm{\bm{v_{ij}}}$ vanishes the second term $\frac{1}{2}\norm{\bm{a_{ij}}}\delta t^{2}$ in Eq.~(\ref{Verlet Position Increment}) can no longer be neglected.

The variable time step scheme is therefore to apply Eq.~(\ref{d Criterion dt}) to every ion pair and to update $\delta t$ at every time step using
\begin{equation}
\delta t = \min_{i,j}\delta t_{i,j}.
\end{equation}

The all pairs computation of this variable time step scheme adds a lot of computation to the already expensive $O(N^{2})$ Coulomb evaluations because it involves three more square roots. Also, it requires more data transfers than the Coulomb force evaluations because it involves not only the particles' positions, but also their speeds and accelerations. This variable time step scheme slows down the code by a factor of $\sim 3$.

\begin{figure}[h]
\begin{center}
\includegraphics[scale=0.8]{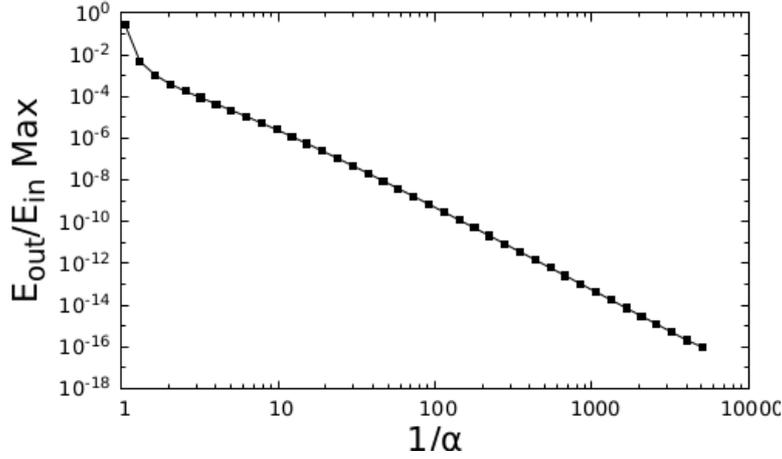}
\caption{Maximum of the ratio of the outgoing energy to the incoming energy $E_{out}/E_{in}$ of the projectile ion after a head-on collision obtained in a 1D model. The maximum is computed over 20000 collisions ranging from 0.1~meV to 1~eV. The ions start \SI{10}{\centi\meter} away.}
\label{Ratio Figure}
\end{center}
\end{figure}

To verify this time step scheme, we simulated head-on 1D collisions of two ions with no trapping force nor laser cooling. We varied the parameter $\alpha$ from $1$ down to $1/5000$. In Fig.~\ref{Ratio Figure} every point shows the maximum value of the ratio of the outgoing energy to the incoming energy of the projectile ion $E_{out}/E_{in}$ over 20000 simulations with collision energies in a geometric progression
from \SI{0.1}{\milli\electronvolt} to \SI{1}{\electronvolt}.
For $\alpha \gtrsim 1$ the time step is too long and the ratio is far from the analytically expected result of zero but as $\alpha$ decreases, the ratio rapidly converges to zero. In Sec.~\ref{section_result}, we use $\alpha=1/100$, which ensures an accurate description of the Coulomb collisions.

As a side note, one may elect to simulate at a constant time step such that Eq.~(\ref{d Criterion 2}) is valid for all ion pairs at all time steps. This may be achieved if the energy of any ion has an upper bound, known in advance, during the duration of a simulation.
One could also upper bound relative velocities and accelerations to twice the maximum velocity and twice the maximum acceleration respectively, and lower bound the distance between two ions by the lowest distance found during the calculation of the Coulomb interaction. That way one could bring down the additional computational cost of the time step calculation to $O(N)$ at the cost of smaller time steps.

\section{$\bar{\textrm{H}}^{+}$ sympathetic cooling}\label{section_result}
\begin{figure}
\begin{center}
\includegraphics[scale=0.7]{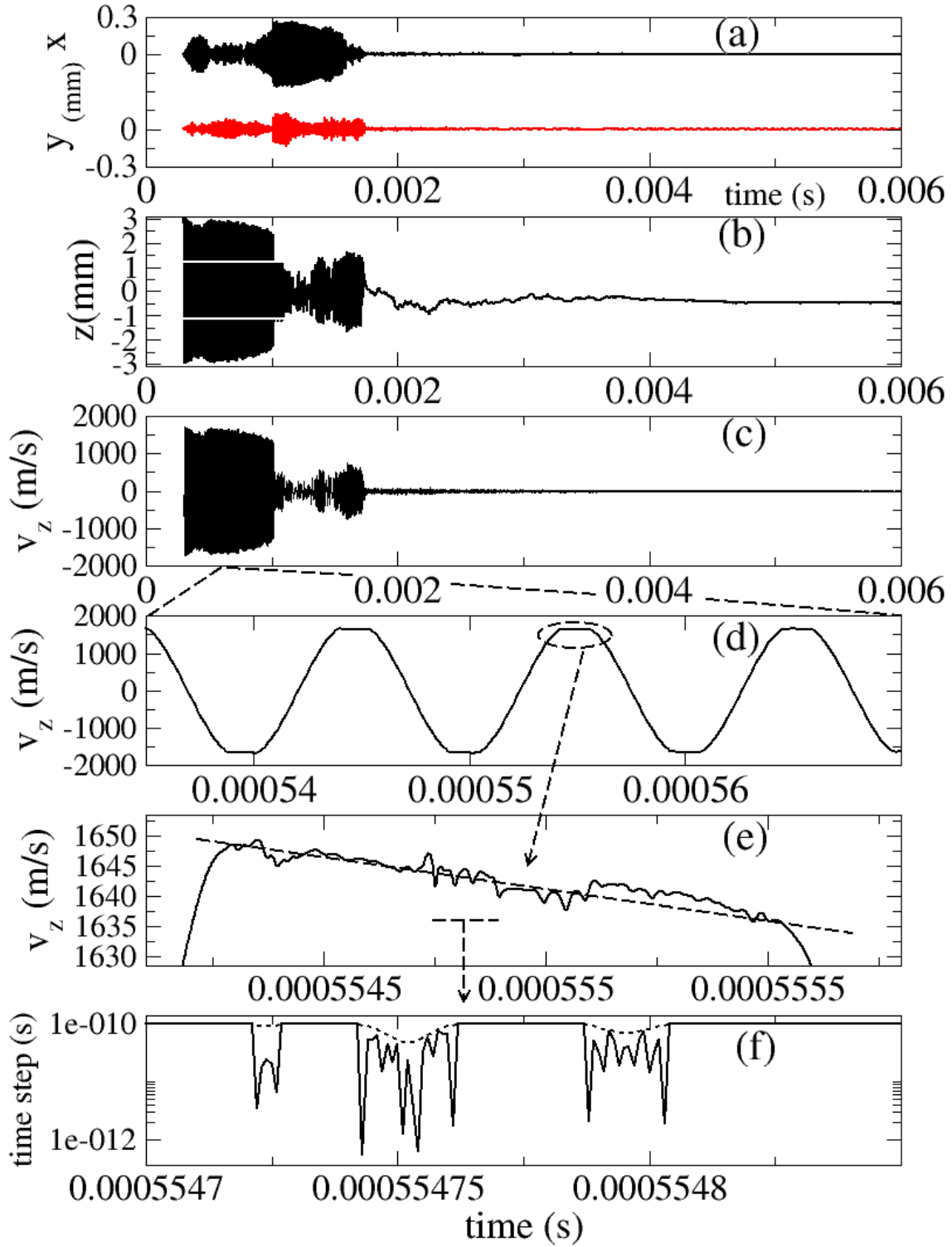}
\caption{Trajectory of an $\bar{\textrm{H}}^+$ ion initially at rest at a distance $z_{0}=$3~mm from the center of the trap containing the Be$^+$/HD$^+$ ion crystal shown in Fig.~\ref{Fig_ion_cloud}. (a)  $x(t)$ and $y(t)$ radial coordinates. (b) Axial position $z(t)$. The white lines indicate the longitudinal Be$^+$/HD$^+$ ion crystal size. (c) Axial velocity $v_z(t)$. (d) Detail of $v_z(t)$ showing oscillations as the $\bar{\textrm{H}}^+$ ion goes back and forth through the ion crystal. (e) Further zooming of $v_z(t)$ showing that the $\bar{\textrm{H}}^+$ ion is slowed down while crossing the ion crystal. (f) Minimum (solid line) and average (dashed line) time step per 10$^{-8}$s interval.
}
\label{Fig_trajectories}
\end{center}
\end{figure}

In this section, we study the capture and sympathetic cooling of an $\bar{\textrm{H}}^+$ ion by a cloud of 500 laser cooled Be$^+$ ions and 200 HD$^+$ ions. The light HD$^+$ ions undergo a tighter radial RF confinement and are located around the trap axis as in Fig.~\ref{Fig_ion_cloud}. The HD$^+$ is sympathetically cooled by the $^9$Be$^+$ and its purpose is to serve as an intermediary of more favorable mass ratio 3, compared to 9, for the sympathetic cooling of the $\bar{\textrm{H}}^{+}$. Indeed it is known from classical mechanics that optimal energy transfer in collisions occurs when the masses are equal. We have performed 5 simulations by varying the random number series used for ion position initialisation and laser interaction. All of them give very similar results and one is detailed in Fig.~\ref{Fig_trajectories}. The simulation first thermalises the 500 Be$^+$ and 200 HD$^+$ ions using laser cooling (not shown in Fig.~\ref{Fig_trajectories}) leading to the ion crystal shown in Fig.\ref{Fig_ion_cloud}.
At $t=$3~$\times$~10$^{-4}$~s, the projectile $\bar{\textrm{H}}^+$ ion is added at rest close to the trap axis
at $x_0=y_0=0$ and $z_0=$ \SI{3}{\milli\meter} with a standard deviation $\Delta z_{0}=\Delta x_{0}=\Delta y_{0}=$ \SI{5}{\micro\meter} corresponding to an initial potential energy of \SI{18.5}{\milli\electronvolt}. Figure~\ref{Fig_trajectories} shows that the projectile ion oscillates back and forth through the ion crystal for several ms before being captured and cooled. When the oscillation amplitude is large, the projectile ion spends most of its time out of the ion crystal where it isn't cooled. Figure~\ref{Fig_trajectories}(d) shows that $v_z(t)$ oscillates with a flat top behaviour corresponding to the Be$^+$/HD$^+$ crystal crossing. This is due to the fact that inside the ion crystal, the total electric field (trapping + Coulomb) is essentially zero such that the projectile ion does not feel any force. Figure~\ref{Fig_trajectories}(e) is a zoom  on the flat top region. It shows that $v_z$ fluctuates due to collisions with the trapped Be$^+$ or HD$^+$ ions. The net effect of the crystal crossing is a slight decrease of the projectile axial velocity that results in projectile capture after many crossings.

Figure~\ref{Fig_trajectories}(f) shows a coarse grain view of the time step evolution with time. The solid and dashed lines correspond to the minimum and average time step over 50~ns time intervals. The minimum time step fluctuates due to Coulomb interactions within the crystal. One can see that when the projectile crosses the crystal, the time step is significantly reduced because the fast ion can come close to the Be$^+$ or HD$^+$ ions.

\begin{figure}
\begin{center}
\includegraphics[scale=0.27]{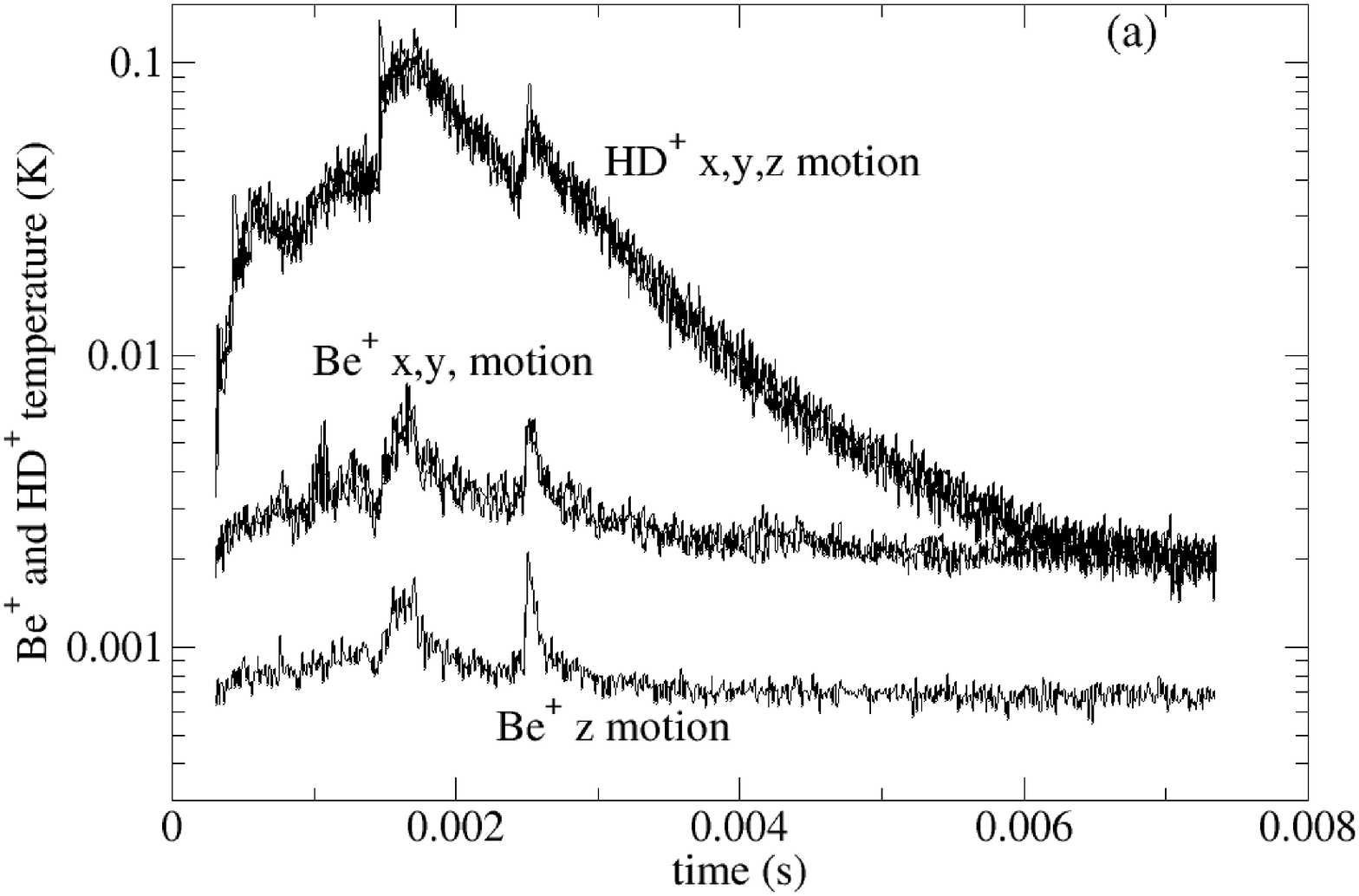}
\includegraphics[scale=0.27]{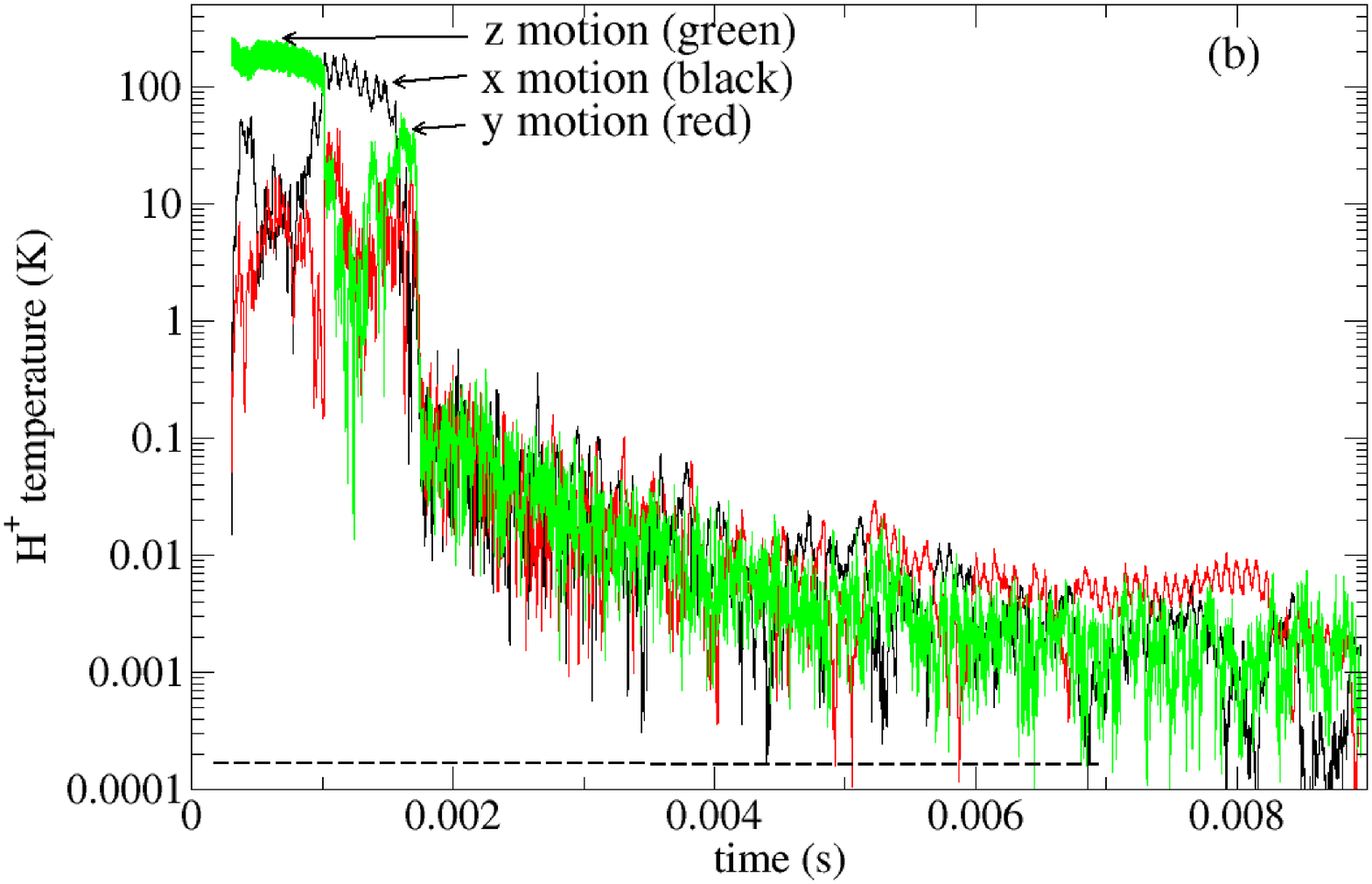}
\caption{(a) For Be$^+$ and HD$^+$: time evolution of the mean macro-motion kinetic energy along the $x$, $y$ and $z$ directions obtained by averaging the velocities over one RF period. (b) Same quantities for the $\bar{\textrm{H}}^+$ projectile ion, but obtained by averaging over 50 RF periods to improve the readability of the curve. The horizontal dashed line indicates the Be$^+$ Doppler limit per degree of freedom, i.e. 0.16~mK.}
\label{Fig_temperature}
\end{center}
\end{figure}

Figure~\ref{Fig_trajectories}(a) shows the trajectories of the projectile ion in the radial plane. Figure~\ref{Fig_temperature}(a) shows the $x$, $y$ and $z$ contributions to the mean macro-motion kinetic energy (expressed in Kelvin) of the Be$^+$ and HD$^+$ ions and Fig.~\ref{Fig_temperature}(b) those of the projectile ion.
One can see that the axial kinetic energy lost by the projectile ion when crossing the ion crystal is partly transferred into radial kinetic energy and also directly to the ion cloud explaining the spikes in the Be$^+$ and HD$^+$ temperatures. This energy is damped by the laser cooling process with a few ms scale leading to cooling of all projectile degrees of freedom and a stable behaviour.
The projectile ion temperature slowly decays from more than 400 K in the z-direction down to the mK regime in less than 10~ms.

\section{Conclusion}
We have shown that simulating ionic interactions incurs a requirement on the time step to properly describe the Coulomb interaction. We have proposed and tested a variable time step scheme to accurately describe the Coulomb interaction while minimising the simulation time.

Using this scheme, we have performed accurate numerical simulations of sympathetic cooling of an $\bar{\textrm{H}}^{+}$ ion by laser cooled Be$^+$ and HD$^+$ ions, showing that sympathetic cooling can be performed in less than 10~ms for a initial kinetic energy of 18.5~meV (more than 400~K). This result is very important to assess the feasibility of the Doppler sympathetic cooling step in the GBAR project. We will pursue these simulations to determine the dependence of the cooling time on the initial energy of the $\bar{\textrm{H}}^{+}$ injected into the laser cooled crystal and on the size of the crystal. We also want to study sympathetic cooling of $\bar{\textrm{H}}^{+}$ by a $\textrm{Be}^{+}$ cloud as the $\textrm{HD}^{+}$ improves cooling but is an extra experimental constraint.

\section{Acknowledgements}

This work was supported by the ANR-13-IS04-0002-01 BESCOOL grant and the COMIQ ITN. J.-Ph. Karr acknowledges Institut Universitaire de France.

\end{document}